\documentclass[twocolumn,10pt]{wlscirep}
\usepackage[utf8]{inputenc}
\usepackage[T1]{fontenc}
\usepackage{bm}
\usepackage{subcaption}
\usepackage{graphicx}
\usepackage{epstopdf}
\usepackage{wrapfig}
\usepackage{array} 
\usepackage{listings}
\usepackage[para,online,flushleft]{threeparttablex}
\usepackage{booktabs,dcolumn}
\usepackage{color}
 \usepackage{textpos}
\usepackage{booktabs}
\usepackage{multirow,bigdelim}
\usepackage{float}
\usepackage{upgreek} 
\usepackage[utf8]{inputenc}
\usepackage{hyperref}
\hypersetup{breaklinks=true,colorlinks=true,linkcolor=blue,citecolor=blue,filecolor=magenta,urlcolor=blue}
\usepackage{xcolor}
\usepackage{mathrsfs}
\usepackage{multirow}
\usepackage{slashed}
\usepackage{slashbox}

\newcommand{\todo}[1]{\color{red}{#1}\color{black} \ }

\title{Investigating the crust of neutron stars with neural-network quantum states}

\author[1]{Bryce Fore}
\author[2]{Jane Kim}
\author[3,4]{Morten Hjorth-Jensen}
\author[1,5,6,*]{Alessandro Lovato}

\affil[1]{Physics Division, Argonne National Laboratory, Argonne, Illinois 60439, USA}
\affil[2]{Institute of Nuclear and Particle Physics and Department of Physics and Astronomy, Ohio University, Athens, Ohio 45701, USA}
\affil[3]{Department of Physics and Astronomy and Facility for Rare Isotope Beams, Michigan State University, East Lansing, Michigan 48824, USA}
\affil[4]{Department of Physics and Center for Computing in Science Education, University of Oslo, N-0316 Oslo, Norway}
\affil[5]{Computational Science Division, Argonne National Laboratory, Argonne, Illinois 60439, USA}
\affil[6]{INFN-TIFPA Trento Institute of Fundamental Physics and Applications, 38123 Trento, Italy}

\affil[*]{lovato@anl.gov}

\begin{abstract}
An accurate description of low-density nuclear matter is crucial for explaining the physics of neutron star crusts. In the density range between approximately 0.01 fm$^{-3}$ and 0.1 fm$^{-3}$, matter transitions from neutron-rich nuclei to various higher-density pasta shapes, before ultimately reaching a uniform liquid. In this work, we introduce a variational Monte Carlo method based on a neural Pfaffian-Jastrow quantum state, which allows us to model the transition from the liquid phase to neutron-rich nuclei microscopically. At low densities, nuclear clusters dynamically emerge from the microscopic interactions among protons and neutrons, which we model based on pionless effective field theory. Our variational Monte Carlo approach represents a significant improvement over the state-of-the-art auxiliary-field diffusion Monte Carlo method, which is severely hindered by the fermion-sign problem in this low-density regime and cannot capture the onset of clusters. In addition to computing the energy per particle of symmetric nuclear matter and pure neutron matter, we analyze an intermediate isospin-asymmetry configuration to elucidate the formation of nuclear clusters. We also provide evidence that the presence of such nuclear clusters influences the amount of protons in the crust compared to protons in beta-equilibrated, neutrino-transparent matter.
\end{abstract}

\begin{document}

\flushbottom
\maketitle
%
%
\thispagestyle{empty}

\section{Introduction}
Low-density nuclear matter found in the crust of neutron stars exhibits a complex and fascinating structure that varies significantly with density. In the outer crust, nucleons are bound to fully ionized nuclei. As density increases within this region, these nuclei become increasingly neutron-rich, and consequently, terrestrial experiments can only directly determine the dominant nuclei species that are present at lower densities. When the neutron drip density is reached, about $2\times10^{-4}$ fm$^{-3}$, the neutron chemical potential becomes too large for nuclei to bind additional neutrons. This leads to the formation of an unbound neutron component, marking the boundary between the outer and inner crust of the star. As the density increases, the proportion of unbound neutrons grows significantly, and at the highest densities, the nuclei undergo phase transitions into various ``pasta'' phases where they deform into non-spherical shapes. Eventually, upon reaching nuclear saturation density, these nuclei coalesce, transitioning into a uniform distribution of nucleons~\cite{Haensel:2007yy,Chamel:2008ca,benhar2023}.

Modeling inner crust material poses significant challenges due to the inherent complexity of clustering phenomena, the emergence of superfluidity, and the existence of both free neutrons and neutron-rich nuclei. Various phenomenological approaches have been developed to address some of these challenges, including the compressible liquid-drop model~\cite{ Baym:1971ax,Baym:1971pw,Ravenhall:1983uh,douchin2001,Newton:2011dw,Gulminelli:2015csa,Lim:2017luh,Carreau:2019zdy,Grams:2021lzx} and self-consistent mean-field models~\cite{Negele:1971vb,Baldo:2006jr,Grill:2011dr}. In the former, the energy is parameterized as a function of global properties, and the nucleons within clusters are treated separately from the free neutrons, all of which are assumed to be uniformly distributed. While this treatment has a low computational cost, it neglects quantum mechanical shell effects, which are critical for determining the equilibrium composition of the crust. In contrast, self-consistent mean-field models are fully quantum mechanical, constructing the many-body state in terms of single-particle or quasi-particle wave functions. However, since they neglect correlations beyond the mean-field approximation, they have limited applicability for low-density nuclear matter, where strong correlations can lead to significant deviations from the mean-field behavior.

Ab initio methods offer a more fundamental approach, as they solve atomic nuclei and neutron-star matter starting from the individual interactions among nucleons~\cite{Hergert:2020bxy}. Most of them, however, suffer from significant limitations when modeling the transition from neutron-rich nuclei to uniform matter~\cite{Day:1967zza,Pandharipande:1979bv,Hagen:2013yba,Carbone:2013eqa}. Their applicability is presently limited to the uniform case---a notable exception in this regard is the lattice effective field theory approach~\cite{Ren:2023ued}. Even continuum diffusion Monte Carlo methods, which are well known for their accuracy across different physical systems, are limited to the uniform phase of neutron-star matter because of the fermion-sign problem~\cite{Schmidt:1999lik,piarulli2020,Lonardoni:2019ypg,Lovato:2022apd}. 

We overcome the limitation of diffusion Monte Carlo approaches by leveraging a variational Monte Carlo method based on neural-network quantum states (NQS)~\cite{Carleo:2017}. Owing to their flexibility, NQS enable an accurate description of the ground-state of quantum many-body systems. This feature is particularly relevant for modeling the different phases of matter comprising the crust of neutron stars, which is inaccessible by conventional ans\"atze. This approach also circumvents the fermion-sign problem, as Monte Carlo techniques are only employed to sample the modulus squared of the wave function, which is positive even for systems of fermions. 

In a recent application~\cite{fore2023}, NQS based on the hidden-nucleon architecture have demonstrated success in modeling the normal and superfluid phases of dilute neutron matter within a unified framework. A more compact and flexible neural Pfaffian-Jastrow architecture, featuring a message-passing neural network to efficiently capture pairing and backflow correlations, was shown to outperform state-of-the-art diffusion Monte Carlo methods in modeling ground-state properties of ultracold Fermi gases~\cite{kim2024}. Chiefly, the neural Pfaffian-Jastrow approach is designed to capture strong pairing effects, which characterize both cold atoms and the matter found in the crust of neutron stars. 

In this work, we put forward a generalization of the neural Pfaffian-Jastrow ansatz and adapt it to nuclear potentials, which strongly depend on the spin-isospin degrees of freedom of the interacting particles. We provide evidence that this ansatz is capable of modeling the different phases matter found in the crust of neutron stars, including the self-emergence of nuclear clusters and their dissolution. In particular, we find dramatically lower energies than those obtained with state-of-the-art auxiliary-field diffusion Monte method in the low-density regime. In addition to ground-state energies, we compute two-body distribution functions and the proton fraction of beta-equilibrated matter. Our findings underscore the viability of utilizing neural Pfaffians to achieve an ab initio description of the crust of neutron stars from the nuclear physics perspective.

\section{Results}

\subsection{Simulation setup}

We model infinite nuclear matter using periodic boundary conditions and $N=28$ nucleons in a cubic simulation cell with side length $L$, where $L$ is determined by the baryon density $n_B=N/L^3$. 
For each density investigated, we run three separate simulations with different proton fractions $x \equiv n_p/n_B = N_p/N$, where $n_p$ is the proton density and $N_p$ is the number of protons in the simulation. Specifically, we run simulations with $N_p =$ 0, 4, and 14 at each density, allowing us to study pure neutron matter (PNM), isospin-symmetric nuclear matter (SNM), and an intermediate isospin-asymmetric case. The simulated nucleons are confined within the cell, with potential contributions from image particles considered up to a distance of $3L/2$. Each nucleon has a continuous three-dimensional position vector and discrete spin and isospin projections on the $z$-axis. As common in quantum Monte Carlo calculations of nuclear systems~\cite{lynn2019,gandolfi2020}, we make the approximation that protons and neutrons have equal masses. 

The dynamics of neutrons and protons is modeled according to a non-relativistic Hamiltonian based on a leading-order pionless effective field theory expansion that contains two- and three-nucleon interactions. This ``essential'' Hamiltonian has been shown to reproduce properties of various nuclei across the nuclear chart, including ground-state energies, radii, and magnetic moments, within few-percent from their experimental values~\cite{schiavilla2021,gnech2023}. It also reproduces well the equation of state (EOS) of dilute neutron matter, as the predicted values agree with highly realistic phenomenological and chiral-EFT interactions~\cite{fore2023}. 
Additionally, we assume the repulsive Coulomb interactions among finite-size protons to be completely screened by the electrons present in the system.

\subsection{Energy per particle}

\begin{figure}
\includegraphics[width=\columnwidth]{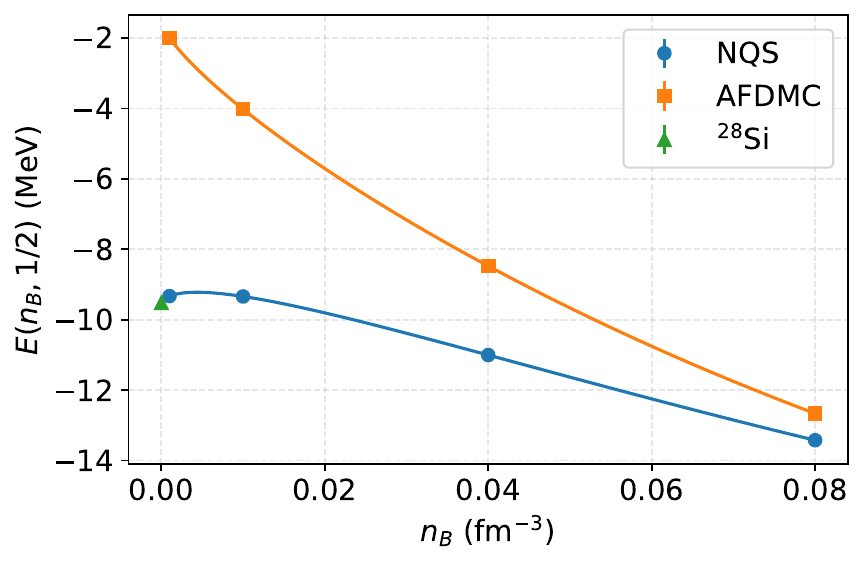}
\caption{Energy per particle of SNM from NQS and AFDMC simulations. The NQS results (blue circles) indicate stronger binding compared to AFDMC (orange squares), attributed to clustering at low densities. Our low-density results are compared to $^{28}$Si with no electromagnetic contribution, which serves as the expected zero-density ground state for a simulation of $N=28$ nucleons with $x=1/2$.}
\label{fig:snm}
\end{figure}

We begin our analysis by comparing the energy per particle of SNM obtained from NQS with that obtained from state-of-the-art auxiliary-field diffusion Monte Carlo (AFDMC), as shown in Figure~\ref{fig:snm}. Details on the AFDMC calculations are provided in the Auxiliary-field diffusion Monte Carlo subsection of the Methods section. For all simulated densities, $n_B = 0.001$ fm$^{-3}$, $0.01$ fm$^{-3}$, $0.04$ fm$^{-3}$, and $0.08$ fm$^{-3}$, the NQS energies are lower than those predicted by AFDMC, especially at the lowest densities. Additionally, the curves exhibit distinctly different characters, with the NQS energies reaching a plateau at lower densities, while AFDMC energies continue to rise. This behavior highlights an important limitation of the AFDMC calculations, as they fail to capture the formation of nuclear clusters. This is ultimately due to the fermion sign problem, which is exacerbated by using a variational ansatz that is designed for the uniform phase. 
Therefore, the zero-density limit of the AFDMC results corresponds to an infinitely dilute gas of protons and neutrons, whose ground-state energy vanishes. As an additional point of reference for the zero-density ground state in the fully clustered system, we present the NQS results for $^{28}$Si with no electromagnetic contribution to the potential, to be consistent with the SNM case. 
The similarity between the NQS energies of $^{28}$Si and SNM at $n_B=0.001$ fm$^{-3}$ suggests that all nucleons in low-density SNM coalesce to form $^{28}$Si. We stress that $^{28}$Si self-emerges as the minimum of the quantum-mechanical energy of the system and is not a priori encoded in the Pfaffian ansatz. 

\begin{figure}
    \includegraphics[width=\columnwidth]{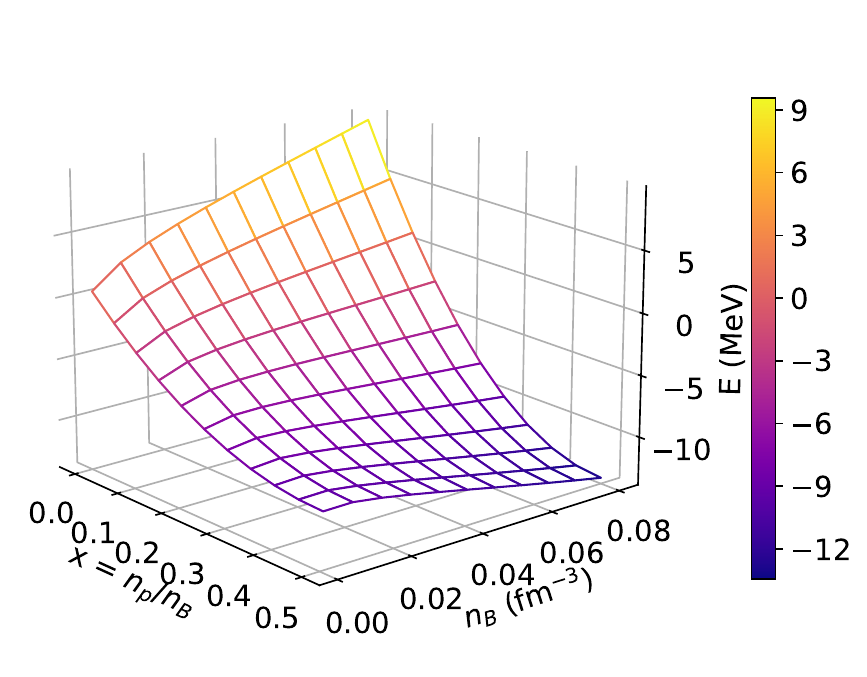}
\caption{Ground-state energy per particle of nuclear matter as a function of baryon density $n_B$ and proton fraction $x$. }
\label{fig:energy_contour}
\end{figure}

Figure~\ref{fig:energy_contour} shows the ground-state energy of nuclear matter as a function of the baryon density $n_B$ and proton fraction $x$. To generate this surface, the NQS energies for SNM and PNM, $E(n_B, 1/2)$ and $E(n_B, 0)$ respectively, were fitted to the following function based on a cluster expansion of the energy\cite{piarulli2020},
\begin{equation}
    E(n_B, x) = a_0 
    + a_{2/3} \left( \frac{n_B}{n_0} \right)^{2/3}
    + a_1 \left( \frac{n_B}{n_0} \right)
    + a_2 \left( \frac{n_B}{n_0} \right)^2.
    \label{eq:density_expansion}
\end{equation}
Here, $n_0=0.16$ fm$^{-3}$ represents the nuclear saturation density and the constant $a_0$ was fixed at zero for PNM. These fits enabled the computation of the symmetry energy $S(n_B)$ as a function of baryon density, defined as the energy difference between PNM and SNM, 
\begin{equation}
    S(n_B)=E(n_B, 0)-E(n_B, 1/2).
    \label{eq:S_def}
\end{equation}
The symmetry energy is an important quantity for characterizing nuclear matter, as it directly affects its neutron richness.
Finally, the energy surface between the two extremes of $x=0$ and $x=1/2$ was approximated through a well-known expansion in neutron richness, $1-2x$, around symmetric matter\cite{lattimer2014}.
The energy computed from this expansion is given by
\begin{equation}
    E(n_B,x)=E(n_B, 1/2)+(1-2x)^2 S(n_B),
    \label{eq:E_S2_expansion}
\end{equation}
where we have already assumed that higher-order terms (which would come in as a fourth power of neutron richness) are negligible, hence the quadratic coefficient is simply the symmetry energy $S(n_B)$ from Eq.~\eqref{eq:S_def}. Note that other expansions have been proposed~\cite{Keller:2024snt}, and found to be more accurate in the low-density regime. We plan to employ them in future NQS calculations of isospin-asymmetric matter, as controlling finite size effects at small $x$ requires using more than $N=28$ nucleons --- see the Appendix on Finite-size effects for additional details. 

From Figure~\ref{fig:energy_contour}, we observe that the energy of nuclear matter decreases as the proton fraction increases, regardless of density. As expected, the change in energy from $x=0$ to $x=1/2$ is more pronounced in the high-density region compared to the low-density region. However, the onset of nuclear clustering makes the energy per particle of SNM at low-density negative by about $-10$ MeV per particle, as shown in Figure~\ref{fig:snm}. Thus,
even at very low densities, the curvature of the energy per particle as a function of $x$ is positive.
Finite-size effects are discussed in more detail in the Finite-size effect corrections subsection of the Methods section.

\subsection{Beta-equilibrated nuclear matter}

Cold nuclear matter at densities found in the inner crust of neutron stars is mostly composed of neutrons, protons, and electrons. Assuming the matter is neutrino transparent, the relative abundance of these particles at a specific density can be determined using a model for the energy as a function of proton fraction, under the well-supported assumptions of charge neutrality, $n_p = n_e$, and beta equilibrium, $\mu_e=\mu_n-\mu_p$~\cite{benhar2023}. We will use this model of beta-equilibrated nuclear matter and our simulation results for SNM and PNM to construct an equation of state for our Hamiltonian. 

\begin{figure}[!b]
\includegraphics[width=\columnwidth]{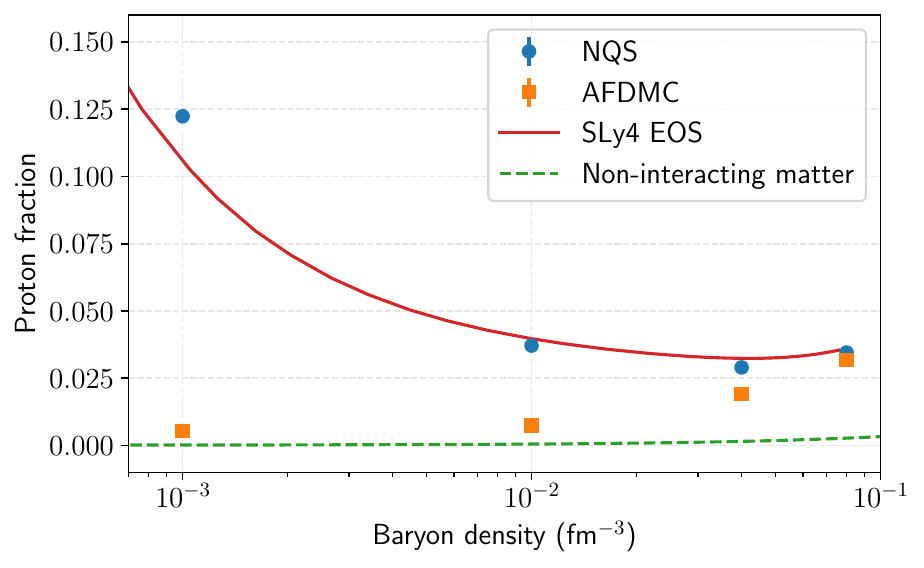}
\caption{Proton fraction in beta-equilibrated matter as a function of baryon density. Blue circles and orange squares corresponds to NQS and AFDMC calculations, respectively. The green dashed line indicates non-interacting matter, and the red solid curve is obtained from a Skyrme parametrization of the equation of state (SLy4 EOS) for the inner crust of neutron stars~\cite{chabanat1998, douchin2001}.}
\label{fig:x_vs_rho}
\end{figure}

The proton fraction in beta-equilibrated matter as a function of baryon density is displayed in Figure~\ref{fig:x_vs_rho}. The NQS predictions align closely with the phenomenological Skryme parameterizations of the equation of state (Sly4 EOS) for the inner crust of neutron stars~\cite{douchin2001,chabanat1998}. Figure~\ref{fig:x_vs_rho} underscores the flexibility of the NQS: by leveraging identical neural network architectures across all densities, we successfully capture the complexity of nuclear structures relevant to the inner crust of neutron stars. In contrast, the AFDMC yields proton fractions that tend towards those of non-interacting matter. This behavior is related to the inability of AFDMC to capture the onset of nuclear clusters at low density, making the presence of protons less energetically favorable compared to more accurate NQS calculations.

Due to the self-formation of nuclei in the low-density region, the NQS energy per particle of SNM obtained for $N=28$ would likely be even lower if we could carry out NQS calculations with $N=40$ particles or more. For example, the experimental energy per particle of $^{28}$Si is $8.45$ MeV, while that of $^{40}$Ca is $8.55$ MeV. These differences would be even larger if Coulomb interactions were turned off. 
On the other hand, the PNM simulations, once corrected for finite-size effects arising from the kinetic energy\cite{fore2023}, are expected to provide results close to the infinite case~\cite{piarulli2020,lovato2022a}. Thus, finite-size effects might underestimate the symmetry energy, yielding too small a proton fraction.

\subsection{Clustering}
The Pfaffian NQS ansatz allows us to capture the self-emergence of nuclei and nuclear clusters from first principles, when their formation corresponds to a lower minimum of the variational energy than the uniform case. To better quantify the impact of clustering, we compute the two-body pair distribution functions, which provide the probability of finding two nucleons with defined isospin at a given distance from each other. The analytical expressions for the distributions are given by
\begin{equation}
    g_\alpha(r) = \frac{1}{2\pi r^2 n_B N} \sum_{i < j}^N \frac{\langle \Psi | \delta (r_{ij}-r) P_\alpha| \Psi\rangle}{\langle \Psi | \Psi \rangle},
\end{equation}
where $P_\alpha$ denotes the projector onto either neutron-neutron ($nn$), proton-neutron ($pn$), or proton-proton ($pp$) pairs. 

\begin{figure}[!hb]
    \includegraphics[width=0.97\columnwidth]{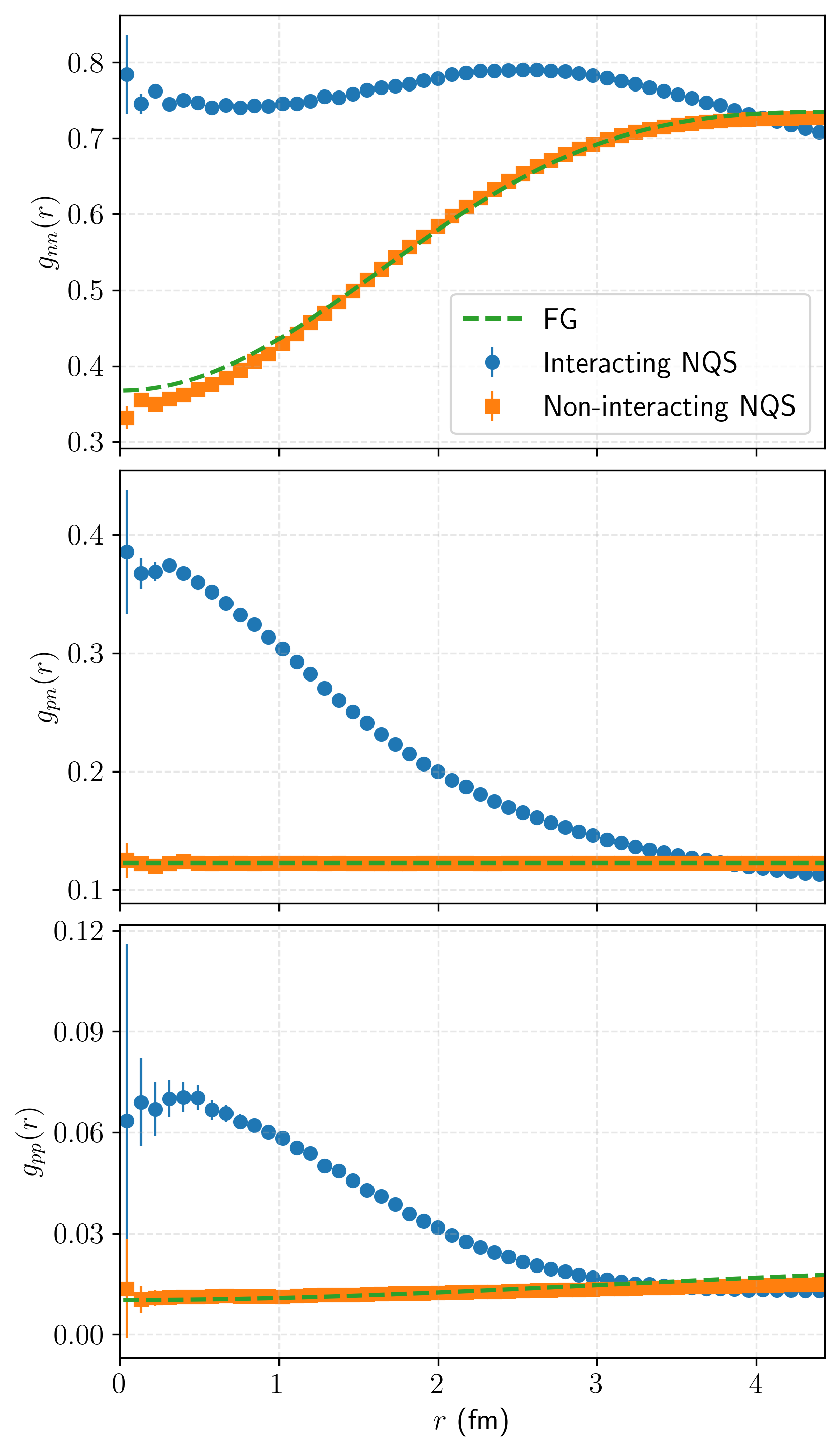}
    
\caption{Pair distribution functions for isospin-asymmetric matter at a density $n_B = 0.04$ fm$^{-3}$. Both the interacting (blue circles) and non-interacting (solid squares) NQS results are obtained simulating 4 protons and 24 neutrons. Two- and three-body potentials in the input Hamiltonian were turned off to generate the non-interacting NQS distributions. The latter agree well with the analytic expressions for the FG in the thermodynamic limit (green dashed curve).}
\label{fig:tbd}
\end{figure}

\begin{figure*}[!htb]
    \centering
    \includegraphics[trim={1cm 1.3cm 1cm 1.3cm},clip, width=\textwidth]{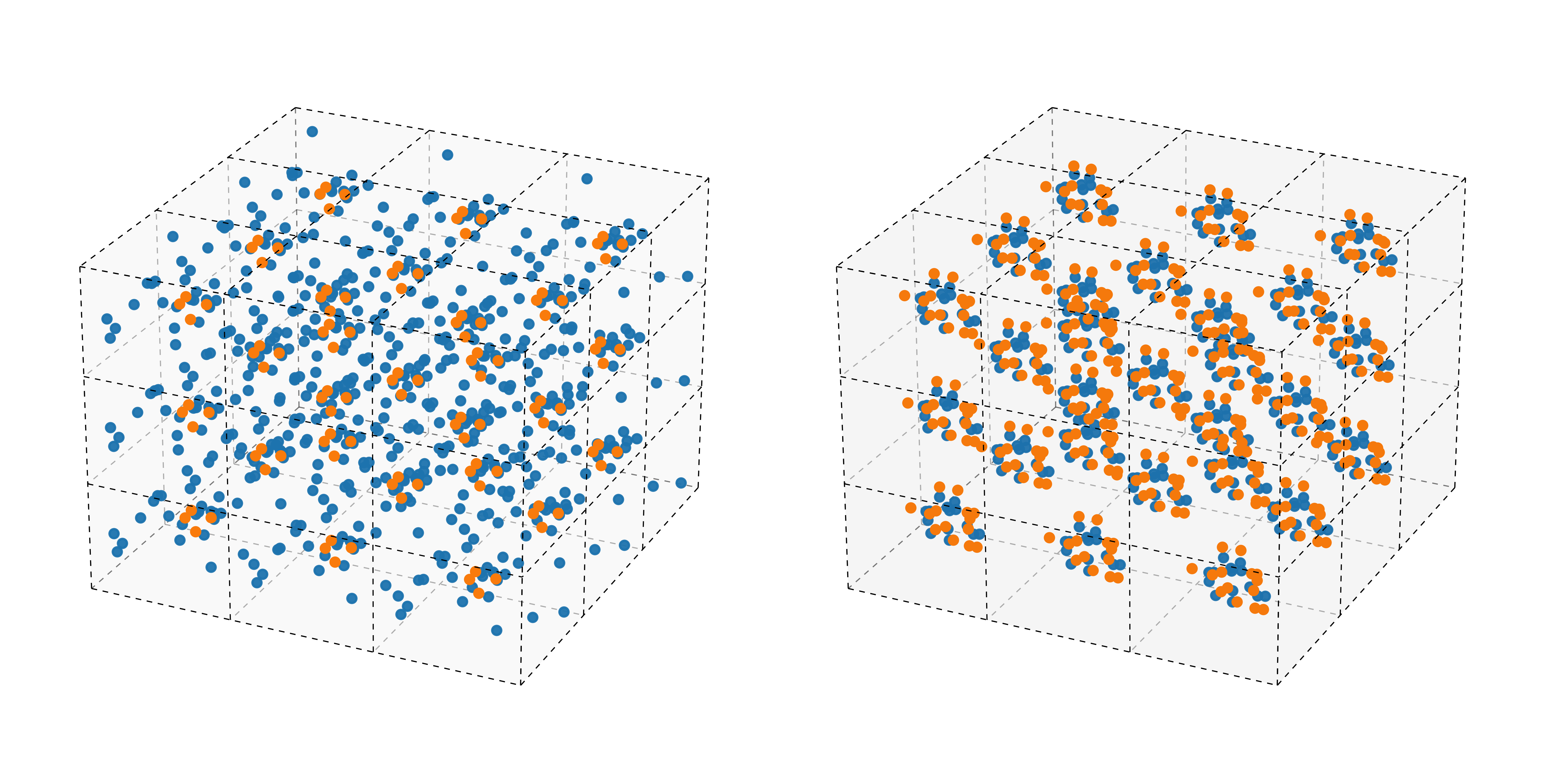}
    \caption{Example configurations at $n_B = $ 0.01 fm$^{-3}$ for two different proton fractions: $x=4/28$ (left) and $x=1/2$ (right). Neutrons are shown in blue and protons in orange. Dashed lines represent the boundaries of the periodic cells. The origin has been shifted to center the clusters of protons within each cell.}
    \label{fig:configurations_3d}
\end{figure*}

Instead of showing $g_\alpha(r)$ for SNM, we find it more informative to focus on an isospin-asymmetric case, specifically $x = 4/28$, as it yields simpler scenarios. The first possibility is the uniform case, in which the distributions should resemble those of the free Fermi Gas (FG), as the input Hamiltonian is relatively soft. A second possibility is the formation of four deuterons, which would yield a peak in the $pn$ distribution but not in the $pp$ one, as the small clusters would float in the simulation box. Finally, if at least one $^4$He cluster is formed, the $pp$ distribution should exhibit a peak at short distances. The two-body distributions displayed in Figure~\ref{fig:tbd} indeed seem to favor the latter case. In fact, compared to the non-interacting NQS case, obtained turning off the two- and three-body potentials in the input Hamiltonian, the distributions show a significant enhancement of pairs at short distances across all channels. For comparison, Figure~\ref{fig:tbd} also displays the FG distributions in the thermodynamic limit, whose expressions can be derived analytically. 

Unfortunately, the analysis of two-body distribution functions alone does not reveal whether heavier nuclei are present, including $^8$Be or more neutron-rich isotopes. To overcome this limitation, we show a three-dimensional snapshot of Monte Carlo configurations for $n_B = $0.01 fm$^{-3}$ and $x = 4/28$ in the left panel of Figure~\ref{fig:configurations_3d}. Clearly, the four protons (represented by solid orange circles) are always close to each other, indicating the presence of $^8$Be. While at least four neutrons (represented by solid blue circles) are consistently close to the protons, there is also evidence of a free neutron component. This behavior is a microscopic indication of the neutron-drip, which is expected to occur in neutron-star matter at these densities. We note that these features are not specific to the Monte Carlo configurations displayed in Figure~\ref{fig:configurations_3d}; rather, analogous structures are evident in all the samples we examined.

In the right panel of Figure~\ref{fig:configurations_3d}, we visualize Monte Carlo configurations for SNM at $n_B = 0.01$ fm$^{-3}$. Consistent with the energy per particle plotted in Figure~\ref{fig:snm}, the $14$ protons and $14$ neutrons coalesce to form $^{28}$Si. As a critical improvement compared to the state-of-the-art AFDMC method, the NQS effectively accommodates the self-emergence formation of a medium-mass nucleus in dilute isospin-symmetric matter. In addition, since the NQS does not require the a priori specification of free neutrons, it provides a more realistic and unbiased representation of the physical system as determined by our Hamiltonian.

\section{Discussion}
In this work, we introduce a highly flexible Pfaffian NQS that leverages backflow transformations through a message-passing neural network to solve the nuclear quantum many-body Schr\"odinger equation with high accuracy. Notably, fermion anti-symmetry, parity, and translation invariance are respected by construction. As a crucial improvement over conventional variational wave functions used in state-of-the-art AFDMC calculations, a single Pfaffian ansatz is able to capture the different phases of matter across a wide range of densities corresponding to the crust of neutron stars. These phases, including extremely neutron-rich nuclei surrounded by superfluid neutrons and a uniform gas of protons and neutrons, are extremely difficult to model within conventional nuclear quantum many-body approaches, with the exception of lattice effective field theory~\cite{Ren:2023ued}.

We demonstrate how the formation of nuclei dramatically decreases the energy of SNM in the low-density region, thereby lowering the value of the symmetry energy. This lowering has a tangible consequence on the proton fraction in beta-equilibrated, neutrino-transparent matter. The Pfaffian NQS yields a proton fraction that agrees well with the one computed using the phenomenological SLy4 functional. The predicted fraction is considerably larger than that obtained from AFDMC calculations in the low-density regime, while it agrees with this method in the uniform case, corresponding to $n_B = 0.08$ fm$^{-3}$ and beyond.

To probe the formation of nuclei heavier than the deuteron, we computed isospin-dependent two-body distribution functions at $n_B = 0.04$ fm$^{-3}$, simulating $4$ protons and $24$ neutrons with periodic-box boundary conditions. The appearance of a peak at short distances in all channels is a clear indication of the presence of nuclei heavier than the deuteron, which is corroborated by Monte Carlo samplings of the many-body wave functions at $n_B = 0.01$ fm$^{-3}$. The latter indicate the formation of neutron-rich Beryllium isotopes, surrounded by free neutrons. The flexibility of the ansatz is such that when simulating SNM at $n_B = 0.01$ fm$^{-3}$, the $14$ protons and $14$ neutrons self-organize to form $^{28}$Si. 

In future calculations, we will leverage leadership-class GPU clusters to simulate a larger number of particles. Reducing finite-size effects will allow us to accurately simulate systems with proton fractions between SNM and PNM, which currently suffer from large systematic uncertainties. Concurrently, we plan to include screened Coulomb repulsion between protons, a fundamental component for investigating signatures of the ``pasta'' phase. A first approach in this direction involves treating the Coulomb lattice of nuclei using the Wigner-Seitz approximation~\cite{Wigner:1934}.

Finally, we are incorporating high-resolution Hamiltonians, both phenomenological and derived from chiral effective field theory, into our variational Monte Carlo method based on NQS. This step is crucial for achieving a unified description of neutron-star matter, from the crust to the inner core.

\section{Method}

\subsection{Neural-network quantum states}
NQS have proven effective in solving the continuous-space Schr\"odinger equation for nuclear many-body systems, ranging from the deuteron to $^{20}$Ne~\cite{Keeble2020,Adams2021,Gnech2021,lovato2022a,Yang:2023rlw,gnech2023}. NQS based on the hidden-nucleon architecture have also successfully modeled both the normal and superfluid phases of dilute neutron matter using a single unified ansatz, demonstrating their flexibility in capturing a wide range of quantum many-body phenomena from first principles. Additionally, a more compact and flexible neural Pfaffian-Jastrow (PJ) architecture, which utilizes a message-passing neural network (MPNN) to efficiently encode pairing and backflow correlations, has been shown to outperform state-of-the-art diffusion Monte Carlo methods in modeling ultracold Fermi gases~\cite{kim2024}. This ansatz is particularly effective at capturing strong pairing effects relevant to both cold atoms and neutron star crust matter.

In this work, we build upon the PJ+MPNN ansatz introduced in Ref.~\cite{kim2024},
\begin{equation}
\begin{split}
& \Psi_{PJ}(X) = \\
& e^{J(X)} \times {\rm pf} 
\begin{bmatrix}
  0 & \phi(\bm{x}_1, \bm{x}_2) & \cdots & \phi(\bm{x}_1, \bm{x}_N) \\
  \phi(\bm{x}_2, \bm{x}_1) & 0 & \cdots & \phi(\bm{x}_2, \bm{x}_N) \\  
   \vdots & \vdots & \ddots & \vdots \\
 \phi(\bm{x}_N, \bm{x}_1) & \phi(\bm{x}_N, \bm{x}_2) & \cdots & 0 \\
  \end{bmatrix}.
\end{split}
\label{eq:PJ}
\end{equation}
To ensure skew-symmetry, the pairing orbital $\phi(\bm{x}_i, \bm{x}_j)$ is defined as 
\begin{equation}
\phi(\bm{x}_i, \bm{x}_j) 
= \nu(\bm{x}_i, \bm{x}_j) 
- \nu(\bm{x}_j, \bm{x}_i),
\label{eq:pair-orb}
\end{equation}
where $\nu(\bm{x}_i, \bm{x}_j)$ is a dense feedforward neural network (FNN). 
Consistent with Ref. \cite{kim2024}, we enforce translational invariance and parity in the ansatz. However, to encompass nuclear potentials, the MPNN is modified to take single-nucleon isospin coordinates through the one-body stream and isospin coordinates for each pair of nucleons through the two-body stream. Additionally, the pairwise spin inputs are modified to use the individual spins of each pair instead of their spin products.

We make a notable improvement to the PJ+MPNN ansatz by allowing the Pfaffian and Jastrow correlator to be complex, while keeping the MPNN real-valued. Generalizing the architecture proposed in Ref.~\cite{bukov2021} for spin systems, we make the Pfaffian complex-valued by feeding the outputs of the MPNN into two separate, real-valued FNNs, which represent the phase and amplitude of $\nu(\bm{x}_i, \bm{x}_j)$. Similarly, the MPNN outputs are fed into two real-valued, permutation-invariant Deep Sets~\cite{Zaheer:2018}, representing the phase and amplitude of the Jastrow correlator $J(X)$.

To help mitigate runaway instabilities~\cite{bukov2021}, the logarithms of the real-valued amplitudes of all the neural networks, generally denoted as $x$, are regularized as
\begin{equation}
x \mapsto a \tanh \left(\frac{x}{a}\right),
\end{equation}
where $a > 0$ is a constant that sets the minimum and maximum allowed values of $x$. For all of our simulations, $a$ was fixed to 8. All the individual FNNs in our ansatz have two hidden layers, each with a width of 16 nodes. The output dimension of the FNNs is either 1 or 16 nodes, depending on whether the FNN maps to a latent space or the output space. The activation function used throughout this ansatz is hyperbolic tangent and the pooling operation is logsumexp. In total, this NQS consists of 10,772 trainable parameters.

\subsection{Optimization}
The optimization of the NQS is performed using a version of the stochastic reconfiguration (SR) algorithm, enhanced with RMSProp-like regularization~\cite{lovato2022a}. For a complex wave function that depends on real parameters, the SR algorithm is modified such that only the real parts of the energy gradient $\bm{g}\equiv \nabla_{\bm{p}}E$ and the quantum Fisher information matrix $S$ are used to update the parameters $\bm{p}$~\cite{bukov2021}, 
\begin{equation}
    \bm{p}_{t} = \bm{p}_{t-1} - \eta \Re(S_{t-1})^{-1} \Re(\bm{g}_{t-1}).
    \label{eq:complex_sr}
\end{equation}
Here, $\eta$ is the learning rate and $t$ is the optimization step. The RMSProp-inspired gradient accumulation is computed as
\begin{equation}
    \bm{v}_t = \beta \bm{v}_{t-1} + (1-\beta) \Re(\bm{g}_{t-1})^2,
\end{equation}
with $\beta = 0.99$. Then, the quantum Fisher information matrix is regularized as
\begin{equation}
    S_t \mapsto S_t + \varepsilon \ \mathrm{diag}(\sqrt{\bm{v}_t} + 10^{-8}),
\end{equation}
where the small constant $\varepsilon > 0$ is typically set to $10^{-3}$. 

All simulations are executed in parallel across 8 NVIDIA A100 GPUs, with a total of 16,000 configuration samples from 800 independent chains. Each optimization step, including Metropolis sampling and observable computation, takes approximately 80 seconds, with convergence generally requiring between 1,000 and 3,000 iterations. For the lowest baryon density examined, $n_B = 0.001$ fm$^{-3}$, transfer learning was employed to improve convergence and stability. Unlike the other densities, which were initialized with random parameters, the optimization for this density utilized parameters previously trained for $n_B = 0.01$ fm$^{-3}$ with the same number of protons and neutrons, significantly reducing the convergence time. 

\subsection{Finite-size effects}
To mitigate finite-size effects, we adjust the NQS energies for a given $n_B$ and $x$ obtained with $N=28$ nucleons by subtracting the free kinetic energy of the finite system and adding the energy of the free FG in the thermodynamic limit~\cite{Gandolfi2014}. These corrections, listed in Table~\ref{tab:finite_correction}, apply to the energies shown in Figure~\ref{fig:snm}, Figure~\ref{fig:energy_contour}, and Table~\ref{tab:all_energies}, as well as those used for calculating the proton fractions in Figure~\ref{fig:x_vs_rho}. 

\begin{table}
\centering
\begin{tabular}{ |c||c|c|c| }
    \hline
    \backslashbox{$n_B$}{$x$} & 0/28 & 4/28 & 14/28\\
    \hline
    \hline
    0.001 & -0.08 & --- & -0.01\\
    0.01 & -0.36 & -0.51 & -0.05\\
    0.04 & -0.91 & -1.29 & -0.13\\
    0.08 & -1.44 & -2.05 & -0.20\\
    \hline
\end{tabular}
\caption{Kinetic energy correction in MeV due to finite-size effects at each density and proton fraction considered. Only SNM and PNM are simulated at the density 0.001 fm$^{-3}$.}
\label{tab:finite_correction}
\end{table}

It should be noted, however, that the above procedure is a rather crude one, as finite-size effects are not limited to the kinetic energy expectation value alone~\cite{Sarsa:2003zu}. Furthermore, leveraging the free FG in the thermodynamic limits is only applicable when clustering is not as prominent as we found. Illuminating in this regard are simulations with low but nonzero proton fractions, as the number of simulated protons may be insufficient to form clusters large enough to significantly lower the energy. To better understand this effect, consider the hypothetical formation of stable nuclei in a sea of free neutrons at rest for a proton fraction of $x=4/28$ in the zero-density limit. Using experimental data for ground-state energies, we can infer that the binding energy per particle for the formation of a $^{8}$Be cluster is $2.02$ MeV, which increases to $2.28$ MeV for $^{16}$O, and further increases to $2.44$ MeV for $^{40}$Ca. This effect becomes even more amplified if the formed nucleus is neutron-rich. For instance, the binding energy is $2.45$ MeV for a $^{12}$Be cluster, $3.02$ MeV for an $^{24}$O cluster, and $3.21$ MeV for a $^{56}$Ca cluster. Therefore, while we can extract useful information to investigate clustering by simulating of $N_p=4$ protons, it is more reliable to employ Eqs.~\eqref{eq:density_expansion} and~\eqref{eq:E_S2_expansion} to interpolate the NQS energies between PNM and SNM.

\subsection{Beta-equilibrated matter solution}
To model neutrino-transparent nuclear matter in beta equilibrium, 
we apply the charge neutrality condition, $n_p = n_e$, and the beta equilibrium condition, $\mu_e = \hat{\mu}$, where $\hat{\mu} \equiv \mu_n - \mu_p$ represents the difference between neutron and proton chemical potentials. Using the well-known expansion of the energy per particle in neutron richness from Eq.~\eqref{eq:E_S2_expansion}, the difference in chemical potentials can expressed in terms of the symmetry energy as
\begin{equation}
    \hat{\mu} = - \frac{\partial E(n_B, x)}{\partial x} = 4 (1-2x) S(n_B).
    \label{eq:mu_hat}
\end{equation}
Next, by calculating the relativistic electron density at zero temperature and applying the charge neutrality condition, we obtain
\begin{equation}
    x\equiv \frac{n_p}{n_B} = \frac{n_e}{n_B} = \frac{(\mu_e^2-m_e^2)^{3/2}}{3\pi^2 n_B}.
\label{eq:x_from_ne}
\end{equation}
Finally, the beta equilibrium condition couples Equations~\eqref{eq:mu_hat} and~\eqref{eq:x_from_ne}, enabling the numerical solution for the proton fraction $x$.

\subsection{Auxiliary-field diffusion Monte Carlo}
\label{sec:AFDMC}

The AFDMC solves the ground-state component of a given Hamiltonian by performing an imaginary-time propagation upon a starting variational ansatz~\cite{Schmidt:1999lik}
\begin{equation}
|\Psi_0\rangle = \lim_{\tau \to \infty} |\Psi(\tau)\rangle = \lim_{\tau \to \infty} e^{-(H-E_V)\tau} |\Psi_V\rangle\, .
\end{equation}
In the above equation, $E_V$ is a normalization constant, which is chosen to be close to the true ground-state energy $E_0$. The variational ansatz is assumed to be of the Slater-Jastrow form: $|\Psi_V\rangle = \hat{F} |\Phi\rangle$. To keep a polynomial cost in the number of nucleons, only linear terms in the spin-isospin correlation operators $\hat{F}$ are included~\cite{Gandolfi2014}. The optimal values of the variational parameters defining the correlation operator are found by employing the same stochastic reconfiguration method used to optimize NQS~\cite{sorella1998}. 

\begin{table}[!b]
\centering
\begin{tabular}{ |c|c||c|c|  }
    \hline
    $x$ & $n_B$ & E (NQS) & E (AFDMC)\\
    \hline\hline
         & 0.001 & $0.702\pm0.001$ & ---\\
    0/28 & 0.01 & $2.510\pm0.001$ & ---\\
         & 0.04 & $6.017\pm0.001$ & ---\\
         & 0.08 & $9.578\pm0.002$ & ---\\
    \hline
         & 0.001 & --- & ---\\
    4/28 & 0.01 & $-1.887\pm0.001$ & ---\\
         & 0.04 & $-1.809\pm0.002$ & ---\\
         & 0.08 & $-1.675\pm0.003$ & ---\\
    \hline
          & 0.001 & $-9.331\pm0.003$ & $-1.997\pm0.046$\\
    14/28 & 0.01 & $-9.341\pm0.004$ & $-4.024\pm0.068$\\
          & 0.04 & $-11.004\pm0.003$ & $-8.476\pm0.034$\\
          & 0.08 & $-13.422\pm0.003$ & $-12.664\pm0.035$\\
    \hline
\end{tabular}
\caption{Values of $E(n_B, x)$, in MeV, from NQS calculations compared with the AFDMC for isospin-symmetric case}
\label{tab:all_energies}
\end{table}

The mean-field part $| \Phi \rangle$ of the starting variational ansatz is the ground state of the uncorrelated Fermi gas, built from a Slater determinant of plane waves. The infinite system is simulated with $N$ nucleons in a cubic periodic box of volume $L^3$. The momentum vectors in this box are discretized as 
\begin{equation}
\mathbf{k} = \frac{2 \pi}{L}\{ n_x, n_y, n_z\}\, ,
\end{equation}
where $n_x$, $n_y$, and $n_z$ are integer numbers. 
In this work, we carry out AFDMC calculations of SNM, simulating $14$ protons and $14$ neutrons as these ``magic'' numbers fill the discretized momentum shells for each species. AFDMC calculations of PNM with $28$ particles in box would require either employing multiple Slater determinants in $|\Phi\rangle$ or twist–averaged boundary conditions~\cite{gandolfi2009a}, which goes beyond the scope of this work. On the other hand, there is excellent agreement between NQS and AFDMC calculations of the energy per particle of PNM carried out with $N=14$ particles within a periodic box~\cite{fore2023}. Thus, to obtain the symmetry energy results labeled as ``AFDMC'' in Fig.~\ref{fig:x_vs_rho}, we use the PNM energies obtained with NQS. 

\subsection{Data}
The ground-state energies of obtained from NQS simulations at different $n_B$ and $x$ are listed in Table~\ref{tab:all_energies}. For the isospin-symmetric case, we also report the corresponding AFDMC values. 

\section{Acknowledgments}
We would like to thank Omar Benhar and Anthea Fantina for illuminating discussions. AL and BF are supported by the U.S. Department of Energy, Office of Science, Office of Nuclear Physics, under contracts DE-AC02-06CH11357, by the 2020 DOE Early Career Award program, and by the NUCLEI SciDAC program. JK is supported by the U.S. Department of Energy, Office of Science, Office of Nuclear Physics, under grant No. DE-SC0021152. MHJ is supported by the U.S. National Science Foundation Grants No. PHY-1404159 and PHY-2013047.

\section{Author Contributions}
B.~F., J.~K., and A.~L. developed the code, designed the Pfaffian-Jastrow ansatz. B.~F. and J.~K. performed the numerical simulations. A.~L. and M.~H.~J. provided overall supervision and guidance throughout the project. All authors contributed significantly to writing the manuscript.

\bibliography{references}

\end{document}